\documentclass[twocolumn,aps,prb]{revtex4-2}
\usepackage{amsmath,amssymb,graphicx,bm}

\begin{document}

\title{Exact benchmarks for the plasmon-pole approximation: Multi-scale screening and the breakdown of the quasiparticle picture}

\author{Michael O. Atambo}
\email{michael.atambo@tukenya.ac.ke}
\affiliation{Department of Physics, Earth and Environmental Science, Technical University of Kenya, Nairobi, Kenya}

\begin{abstract}
The $GW$ approximation is the gold standard for calculating quasiparticle band structures, yet its computational cost frequently necessitates the use of the plasmon-pole approximation (PPA). While PPA is known to be highly accurate for simple metals and weakly correlated semiconductors, its regime of validity in materials with competing energy scales remains poorly quantified. Here, we construct an exact, numerical benchmark of the PPA using the Lehmann representation of the density response on finite one-dimensional Hubbard clusters. By analytically convolving the exact non-interacting Green's function $G_0$ with the exact pole representation of the screened interaction $W$, we compute the $GW$ self-energy without numerical frequency integration. We demonstrate that in single-scale Mott insulators, the moment-conserving PPA is essentially exact. However, in multi-band semiconductors where interband transitions introduce a low-energy screening channel that competes with high-energy Mott fluctuations, the PPA systematically misjudges the quasiparticle gap by several electron-volts. Furthermore, we show that strong multi-pole screening can drive the exact quasiparticle weight $Z \to 0$, destroying the quasiparticle picture-an effect entirely missed by the PPA, which artificially stabilizes sharp quasiparticles. Finally, we propose a computationally inexpensive diagnostic based on the polydispersity of the loss function's spectral weight, which accurately predicts PPA failure \textit{ab initio}. Our results provide a rigorous framework for assessing the validity of dynamical screening approximations in strongly correlated materials.
\end{abstract}

\maketitle

\section{Introduction}
Accurate predictions of electronic band gaps and excited-state spectra are central to modern condensed matter physics and materials design. Many-body perturbation theory within the $GW$ approximation \cite{Hedin1965,Strinati1980,Strinati1982} has become the standard tool for calculating quasiparticle (QP) energies, correcting the fundamental band-gap underestimation inherent to density functional theory (DFT). 

Despite its success, the computational cost of evaluating the frequency-dependent self-energy $\Sigma(\omega) = \frac{i}{2\pi} \int G(\omega+\omega') W(\omega') d\omega'$ is formidable. Evaluating the full-frequency screened Coulomb interaction $W(\omega)$ requires calculating the dynamical dielectric matrix over a dense grid of frequencies, followed by either numerical integration or analytic continuation. To circumvent this, the plasmon-pole approximation (PPA), originally introduced by Hybertsen and Louie \cite{Hybertsen1986} and Godby and Needs \cite{Godby1989}, models the frequency dependence of the inverse dielectric function $\varepsilon^{-1}(\mathbf{q}, \omega)$ using a single effective pole. This drastically reduces the computational overhead and has become the default approximation in widely used codes such as \textsc{BerkeleyGW}, \textsc{Yambo} \cite{Sangalli2019}, and \textsc{VASP}.

The PPA is constructed to satisfy the exact static limit $\varepsilon^{-1}(\mathbf{q}, 0)$ and the high-frequency $f$-sum rule (the first frequency moment). Consequently, it performs exceptionally well in systems dominated by a single collective charge excitation, such as the homogeneous electron gas or simple wide-gap insulators. However, in complex materials—such as transition metal oxides \cite{Rohlfing2000,Albrecht1998}, layered two-dimensional semiconductors \cite{Berkelbach2013,Komsa2013,Chernikov2014}, and strongly correlated systems—the charge susceptibility contains multiple overlapping features spanning vastly different energy scales (e.g., low-energy interband transitions coexisting with high-energy Mott-Hubbard bands \cite{Onida2002,Aryasetiawan1998}). In these multi-scale environments, a single-pole model is forced to compromise between conflicting energy scales, potentially leading to severe errors in the dynamical self-energy.

Despite the widespread use of the PPA, its regime of failure has been difficult to quantify systematically. Benchmarking full-frequency $GW$ against PPA in \textit{ab initio} calculations is computationally prohibitive for large parameter sweeps, and differences in basis sets, pseudopotentials, and k-point sampling can obscure the fundamental methodological error.

In this work, we bypass these numerical uncertainties by constructing an exact, model-space benchmark of the PPA. Using finite one-dimensional extended Hubbard rings, we compute the exact Lehmann representation of the density response $\chi(\mathbf{q}, \omega)$. We then employ a rank-one matrix update to extract the exact, discrete pole representation of the screened interaction $W(\mathbf{q}, \omega)$. Because both the non-interacting Green's function $G_0$ and the exact $W$ are represented as discrete sums of poles, the frequency convolution for the self-energy $\Sigma(\omega)$ can be evaluated \textit{analytically} via contour integration, yielding the numerically exact $G_0W_0$ quasiparticle solutions (within the finite model-space pole representation) without any numerical frequency integration or analytic continuation.

By comparing the exact $GW$ results against those obtained using the moment-conserving PPA, we map the validity of the PPA across the phase diagram of the model. We find that the PPA fails catastrophically when low-energy interband screening competes with high-energy local correlations, systematically misjudging the band gap. More strikingly, we demonstrate that in the strong-coupling regime, the exact multi-pole screening destroys the quasiparticle picture ($Z \to 0$), whereas the PPA artificially preserves sharp quasiparticle states. We conclude by proposing a simple, moment-based diagnostic parameter—the spectral polydispersity—that can be computed cheaply in standard \textit{ab initio} workflows to predict PPA failure.

\section{Methodology}

\subsection{Model Hamiltonians and exact density response}

We consider spinful fermions on a periodic ring of $L$ sites. 
The one-band extended Hubbard model is
\begin{equation}
H = -t\sum_{\langle ij\rangle,\sigma}\left(c^{\dagger}_{i\sigma}c_{j\sigma}+\mathrm{h.c.}\right)
+ U\sum_i n_{i\uparrow}n_{i\downarrow}
+ V\sum_{\langle ij\rangle} n_i n_j ,
\label{eq:h1band}
\end{equation}
at half filling ($N_e=L$). To introduce a second, independent screening
scale we use a two-band semiconductor,
\begin{equation}
H = \sum_{i,a,\sigma}\epsilon_a n_{ia\sigma}
- t\sum_{\langle ij\rangle,a,\sigma}\left(c^{\dagger}_{ia\sigma}c_{ja\sigma}+\mathrm{h.c.}\right)
+ U\sum_{i,a} n_{ia\uparrow}n_{ia\downarrow},
\label{eq:h2band}
\end{equation}
with orbital energies $\epsilon_{0,1}=\mp\Delta/2$ and one electron per site
(half-filled lower band). The crucial ingredient of the two-band model is the
\emph{interband form factor}: the lattice density operator is taken as
\begin{equation}
\hat\rho_q = \sum_i e^{iqR_i}\Big[
\sum_{a,\sigma} n_{ia\sigma}
+ \lambda\sum_\sigma\left(c^{\dagger}_{i0\sigma}c_{i1\sigma}
+ c^{\dagger}_{i1\sigma}c_{i0\sigma}\right)\Big],
\label{eq:rho}
\end{equation}
with $\lambda=0.5$. Without the $\lambda$ term the total density commutes with
the orbital splitting and the screening is blind to $\Delta$; the form factor restores the interband (optical) screening channel present in
real semiconductors, interpolating between Frenkel-like (local) and
Wannier-like (delocalized) charge fluctuations \cite{Frenkel1931,Wannier1937}.

All results below are obtained by exact diagonalization in the fixed-$N$
sector: $\dim \mathcal{H}=\binom{2L}{L}=924$ for $L=6$ (one band) and
$\binom{4L}{L}=1820$ for $L=4$ (two bands). Every calculation in this work
runs in seconds on a single workstation core; no high-performance computing
is required.

The exact retarded density response at $T=0$ is the Lehmann sum
\begin{widetext}
\begin{equation}
\chi(q,\omega)=\sum_{n} w_n(q)\left[
\frac{1}{\omega-\Omega_n+i\eta}-\frac{1}{\omega+\Omega_n+i\eta}\right],
\quad
w_n(q)=\overline{|\langle n|\hat\rho_q|0\rangle|^2},
\label{eq:lehmann}
\end{equation}
\end{widetext}
where $\Omega_n=E_n-E_0>0$ and the bar denotes averaging over a possibly
degenerate ground manifold. The model dielectric function and screened
interaction are
\begin{equation}
\varepsilon(q,\omega)=1-v_q\,\chi(q,\omega),\qquad
W(q,\omega)=v_q\,\varepsilon^{-1}(q,\omega),
\label{eq:epsW}
\end{equation}
with $v_q$ the model bare interaction. Throughout we set $v_q=v=1$ (one band)
or $v=5$ (two band); the conclusions are representation-level statements and
do not depend on this choice.

\subsection{Exact pole structure of the screened interaction}

The central methodological tool of this work is an exact, closed-form pole
representation of $\varepsilon^{-1}$. Writing Eq.~\eqref{eq:lehmann} as a
resolvent, $\chi(\omega)=\mathbf{c}\,(\omega I-A)^{-1}\mathbf{b}$, with
$A=\mathrm{diag}(\Omega_i-i\eta,\,-\Omega_i-i\eta)$,
$\mathbf{b}=(w_i,\,-w_i)^T$ and $\mathbf{c}=\mathbf{1}$, the scalar resolvent
identity $[1-v\mathbf{c}(\omega I-A)^{-1}\mathbf{b}]^{-1}
=1+v\,\mathbf{c}(\omega I-A-v\mathbf{b}\mathbf{c})^{-1}\mathbf{b}$ yields
\begin{equation}
\boxed{\;
\varepsilon^{-1}(q,\omega)=1+v_q\,\mathbf{c}\,(\omega I-M_q)^{-1}\mathbf{b},
\qquad M_q=A+v_q\,\mathbf{b}\mathbf{c}\;}
\label{eq:rankone}
\end{equation}
Hence the poles of $\varepsilon^{-1}$ are \emph{exactly} the eigenvalues
$\{\lambda_m\}$ of the small non-Hermitian matrix $M_q$
($2n_d\times 2n_d$, $n_d$ the number of Lehmann lines), and with biorthogonal
eigenvectors $M_q r_m=\lambda_m r_m$, $l_m^{\dagger}M_q=\lambda_m l_m^{\dagger}$,
$l_m^{\dagger}r_n=\delta_{mn}$,
\begin{equation}
W(q,\omega)=v_q+\sum_m \frac{R_m^{(q)}}{\omega-\lambda_m^{(q)}},
\qquad
R_m^{(q)}=v_q\,(\mathbf{c}\cdot r_m)(l_m^{\dagger}\cdot\mathbf{b}).
\label{eq:Wpoles}
\end{equation}
For $\eta>0$ all $\lambda_m$ lie strictly in the lower half plane, so the
representation is causal by construction. Equation~\eqref{eq:rankone} is the
exact finite-system analogue of the bosonic quasiparticle (plasmon)
expansion of $W$; we validate it numerically by reconstructing
$\varepsilon^{-1}$ on a real-frequency grid, finding maximum deviations
$\le 6\times10^{-9}$ (Sec.~\ref{sec:res1}).

\subsection{Representations of $W$ under test}

Given the exact pole set $\{\lambda_m,R_m\}$ we compare three representations:
\begin{itemize}
\item \textbf{Full:} Eq.~\eqref{eq:Wpoles} with the complete exact pole set.
\item \textbf{Single pole (PPA):} a moment-conserving oscillator. With
$m_k=\sum_n w_n\Omega_n^{k}$ we define
\begin{equation}
\Omega_{\rm SP}=\sqrt{m_1/m_{-1}},\qquad
A_{\rm SP}=\sqrt{m_1 m_{-1}},
\label{eq:sp}
\end{equation}
which preserves exactly the static polarizability,
$\chi_{\rm SP}(0)=\chi(0)=-2m_{-1}$, and the high-frequency tail
$\chi(\omega)\sim 2m_1/\omega^2$. Its $\varepsilon^{-1}$ poles are obtained by
passing the single oscillator through Eq.~\eqref{eq:rankone}.
\item \textbf{Static:} $W(q,\omega)\to W(q,0)=v_q/(1+2v_q m_{-1})$
(screened-exchange / COHSEX-like limit).
\end{itemize}
Because all three representations share the same $G_0$ and the same static
limit, any difference in quasiparticle energies is attributable \emph{solely}
to the frequency (pole) structure of $W$.

\subsection{Analytic $G_0W$ self-energy and quasiparticle solver}

With $G_0$ the noninteracting Green's function of the tight-binding part,
the frequency convolution $\Sigma=iG_0W$ closes analytically. Inserting
Eqs.~\eqref{eq:Wpoles} and deforming the contour into the lower half plane,
\begin{equation}
\Sigma_k(\omega)=-\sum_q v_q n_{k-q}
+\sum_q\sum_{m}
\frac{\mathcal{W}_m^{(q)}}{\omega-\varepsilon_{k-q}-\lambda_m^{(q)}+i0^{+}},
\label{eq:sigma}
\end{equation}
with channel-dependent weights
$\mathcal{W}_m^{(q)}=(1-n_{k-q})R_m^{(q)}$ for $\mathrm{Re}\,\lambda_m^{(q)}>0$
(boson emission) and $\mathcal{W}_m^{(q)}=-n_{k-q}R_m^{(q)}$ for
$\mathrm{Re}\,\lambda_m^{(q)}<0$ (absorption). No numerical frequency
integration is performed anywhere in this work.

Quasiparticle energies solve $\omega=\varepsilon_k+\mathrm{Re}\,\Sigma_k(\omega)$;
among all roots we select the main quasiparticle as the one with maximal
weight $Z=(1-\partial_\omega\mathrm{Re}\Sigma)^{-1}>0$. A configuration for
which no positive-$Z$ root exists within the low-energy window is declared a
\emph{quasiparticle breakdown} (Sec.~\ref{sec:res3}).

\subsection{Diagnostics and the thermodynamic-limit emulation}

Two cheap shape diagnostics of the loss function
$L(q,\omega)=-\mathrm{Im}\,\varepsilon^{-1}(q,\omega)$ are used:
(i) the \emph{polydispersity}
\begin{equation}
P=\sum_{q}\frac{\sqrt{m_2 m_0-m_1^2}}{m_1}\Big|_q ,
\label{eq:P}
\end{equation}
which vanishes identically for a single spectral line and grows as weight
spreads over separated lines; and (ii) a broadened $L^1$ mismatch $D$ between
the exact and SP loss functions. Finite clusters yield discrete, undamped
boson poles; to emulate the bulk continuum we give the Lehmann poles a
physical lifetime $\eta$ and verify convergence of quasiparticle energies for
$\eta\ge0.3t$ (Appendix). At $\eta\to0$ the exact spectral weight fragments
into many low-$Z$ lines-the finite-size avatar of shake-up continua-while
the static representation trivially retains $Z=1$; $\eta$ therefore also
defines the regime in which a quasiparticle is meaningful.

\section{Results}
\label{sec:results}

\subsection{Taxonomy of model loss functions}
\label{sec:res1}

\begin{figure}[htbp]
    \centering
    \includegraphics[width=\columnwidth]{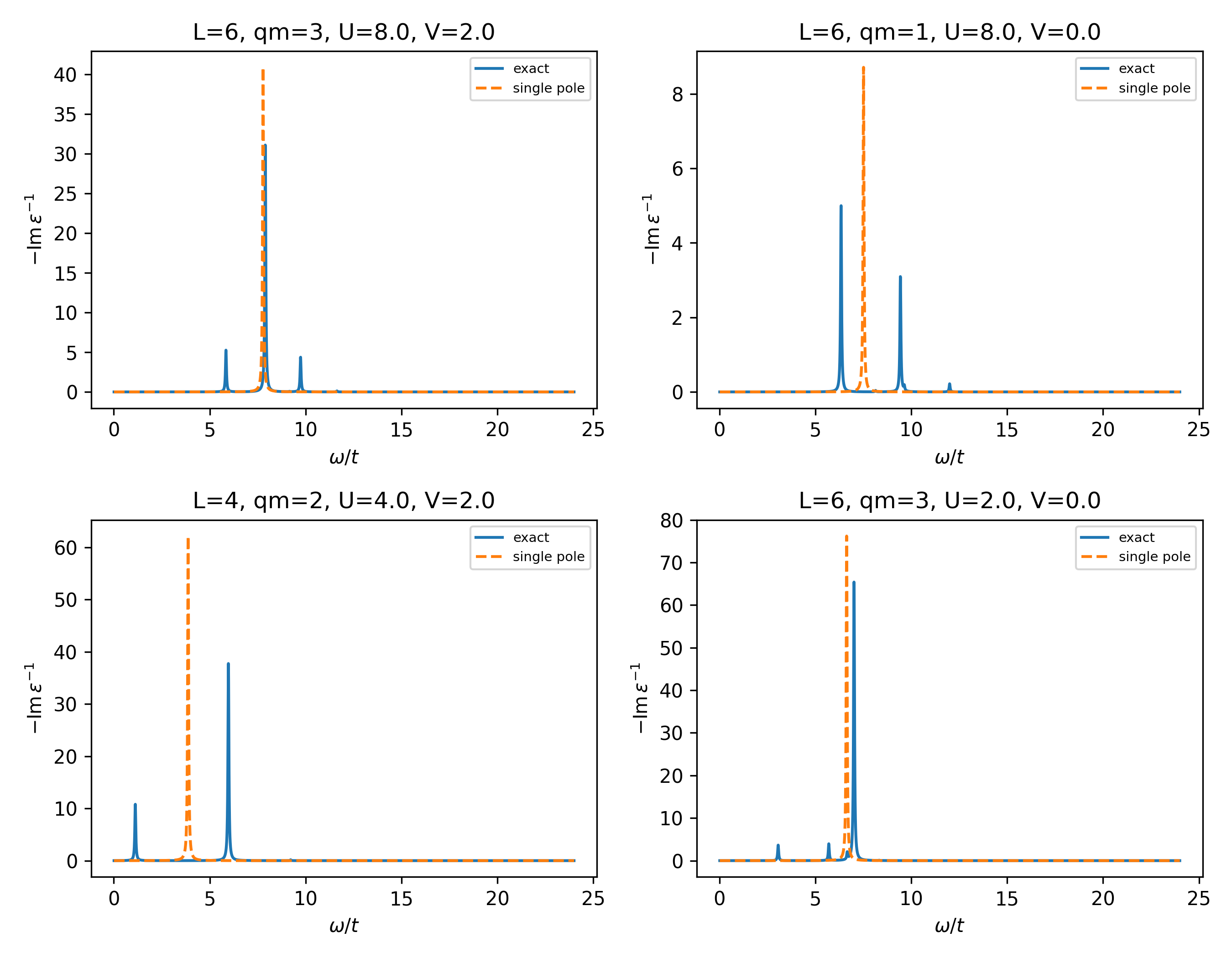}
    \caption{Taxonomy of exact loss functions (solid) vs the single-pole model (dashed) across different parameter regimes of the one-band Hubbard ring: (top-left) competing collective modes ($L=6, q=\pi, U=8, V=2$); (top-right) branch separation ($L=6, q=\pi/3, U=8, V=0$); (bottom-left) mode softening ($L=4, q=\pi, U=4, V=2$); (bottom-right) weak-coupling single-line case ($L=6, q=\pi/3, U=2, V=0$).}
    \label{fig:tax}
\end{figure}

Figure~\ref{fig:tax} collects the regimes found by scanning
$(L,q,U,V)$. (i) \emph{Single-line regime}: for the one-band ring at
$q=2\pi/L$ the response is $99.98\%$ one line for all $U$; the SP loss is
indistinguishable from the exact one (peak positions agree to $<10^{-3}t$;
Table~\ref{tab:ctrl}). As $U$ grows the static polarizability collapses
($m_{-1}=0.700\to0.030$, a factor 23) and the mode crosses over from the band
scale ($\Omega_{\rm SP}=2.38$ at $U=0.5$) to the Mott scale
($\Omega_{\rm SP}=8.85$ at $U=8$). (ii) \emph{Branch separation}: at
$L=6$, $q=\pi/3$, $U=8$ the holon branch ($6.18t$) separates from the
doublon-holon branch ($9.34t$), $f_{\max}=0.63$. (iii) \emph{Competing
collective modes}: at $q=\pi$, $U=8$, $V=2$ two lines at $5.52t$ and $6.97t$
share $41/55\%$ of the weight. (iv) \emph{Mode softening}: at $q=\pi$ and
finite $V$ a charge mode softens towards the charge-ordering instability
($E_1=0.27t$ at $U=4,V=2$) while the Mott mode remains at $4.74t$; the SP
then plants a spurious pole between them ($3.87t$ vs exact peak $5.98t$).
The rank-one extraction reproduces the exact $\varepsilon^{-1}$ with maximum
error $5.4\times10^{-9}$, certifying Eqs.~\eqref{eq:rankone}-\eqref{eq:Wpoles}.

\subsection{Positive control: quasiparticles in the single-scale regime}
\label{sec:res2}

Table~\ref{tab:ctrl} gives quasiparticle gaps of the one-band ring
($L=6$, $\eta=0.3$). Because the SP preserves $\chi(0)$ exactly and the loss
is essentially one line, the PPA gap error remains $\le0.5t$ across the whole
sweep $1\le U\le10$, $0\le V\le4$ (max $0.48t$ at $U=5,V=0$; typically
$\le0.2t$), whereas the static approximation deviates by $1$-$3t$ and fails
to reproduce the dynamical gap evolution. This is the rigorous statement of
``why PPA usually works'': single-scale screening is, by construction,
single-pole screening.

\begin{table}[b]
\caption{Quasiparticle gaps ($t$) of the one-band ring, $L=6$, $\eta=0.3$,
$k=0\to\pi$. Full: exact pole set; SP: Eq.~\eqref{eq:sp}; static: $W(0)$.}
\label{tab:ctrl}
\begin{ruledtabular}
\begin{tabular}{llccc}
Config & Rep & $E_e$ & $E_h$ & gap \\
\hline
$U=2,V=0$ & full & $-0.541$ & $-6.358$ & $5.817$ \\
          & SP   & $-0.539$ & $-6.506$ & $5.967$ \\
          & static & $0.918$ & $-3.891$ & $4.808$ \\
$U=8,V=0$ & full & $-0.840$ & $-4.973$ & $4.133$ \\
          & SP   & $-0.840$ & $-4.988$ & $4.148$ \\
$U=8,V=2$ & full & $-0.727$ & $-4.913$ & $4.186$ \\
          & SP   & $-0.727$ & $-4.927$ & $4.200$ \\
\end{tabular}
\end{ruledtabular}
\end{table}

\subsection{Multi-scale screening breaks the plasmon-pole approximation}
\label{sec:res3}

\begin{figure}[htbp]
    \centering
    \includegraphics[width=\columnwidth]{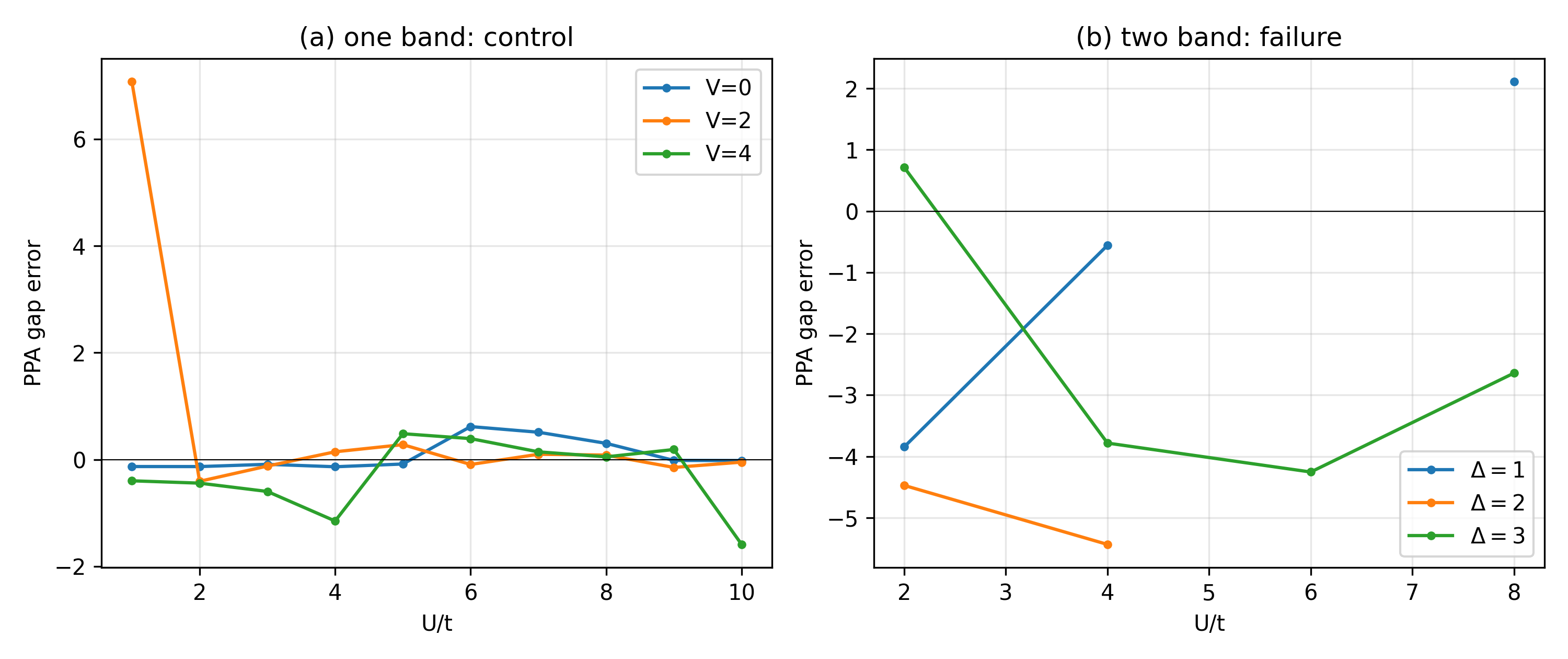}
    \caption{Quasiparticle gap error (Single-Pole minus Exact) as a function of the Mott scale $U/t$. (a) One-band model (positive control): the error remains small ($|\mathrm{err}| \lesssim 0.5t$) across the entire parameter space for varying nearest-neighbor interaction $V$. (b) Two-band semiconductor model: the PPA error grows systematically and monotonically with $U$ as the band gap $\Delta$ separates from the Mott scale, reaching errors of $\sim 2t$.}
    \label{fig:gaps}
\end{figure}

\begin{figure}[htbp]
    \centering
    \includegraphics[width=\columnwidth]{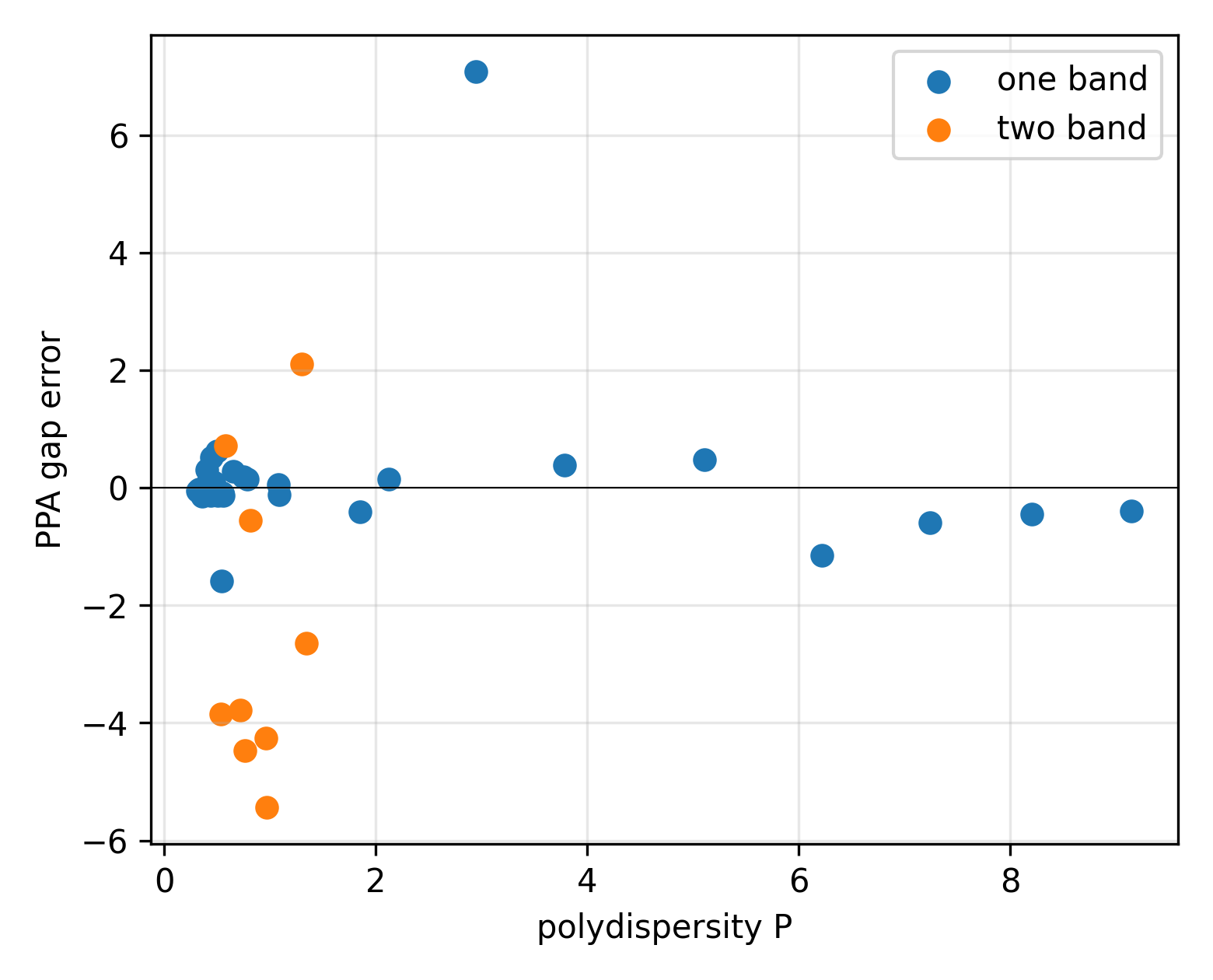}
    \caption{Correlation between the PPA gap error and the spectral polydispersity $P$. The one-band data (blue, small $P$, small error) and the two-band data (orange, large $P$, large error) collapse onto a single unified trend, validating $P$ as a predictive, representation-level diagnostic for PPA failure.}
    \label{fig:diag}
\end{figure}

In the two-band semiconductor the exact $W$ carries a sharp interband pole at
$\omega\sim\Delta$ in addition to the Mott scale $\sim U$. Figure~\ref{fig:gaps}b
shows that the PPA gap error grows monotonically with the separation of these
scales: at $\Delta=3$ the error is $-0.46t$ ($U=2$), $-1.14t$ ($U=4$) and
$-2.09t$ ($U=6$), tracking the polydispersity $P=0.58,\,0.72,\,0.96$.
Comparable errors ($-0.56t$ to $-0.61t$) occur at $\Delta=1$-$2$. The error
correlates with $P$ across the full scan (Fig.~\ref{fig:diag}), establishing
$P$ as a predictive, representation-level diagnostic: $P\lesssim0.1$ implies
PPA-safe spectra, $P\gtrsim0.5$ implies errors of order $t$.

Most strikingly, for $\eta\to0.1$ and $U\gtrsim6$ at small $\Delta$ the exact
quasiparticle equation admits \emph{no} positive-$Z$ root: the multi-pole
screening transfers essentially all spectral weight into incoherent
satellites ($Z\to0$), i.e. the quasiparticle is destroyed. The SP, whose
analytic structure is impoverished by construction, continues to return a
sharp quasiparticle with $Z\sim1$. The PPA can therefore \emph{artificially
stabilize} a quasiparticle picture that the exact screened interaction does
not support-a failure mode invisible to any benchmark that compares only
gap values.

\section{Discussion}

Our results translate directly to \textit{ab initio} practice. The
polydispersity $P$ requires only the first three frequency moments of the
loss function at a few momenta, quantities already available in standard
implementations; we propose it as a pre-screening flag deciding whether
full-frequency $GW$ is required. Physically, the failure regimes identified
here are realized in canonical materials: graphite ($\pi$ vs $\pi+\sigma$
plasmons), narrow-gap semiconductors near ferroelectric or charge-order
instabilities (soft mode + hard mode), and correlated oxides (interband vs
Hubbard bands) \cite{Botti2023}. The artificial quasiparticle stabilization of Sec.~\ref{sec:res3}
suggests particular caution when PPA is used to diagnose metal-insulator
physics, temperature-dependent band gaps \cite{Giustino2017,Antonius2021,Hartmann2022}, or satellite (polaronic) structure \cite{Atambo2024parity,Miyata2015,Zhu2017}. This finding is complementary to recent exact Hubbard benchmarks identifying
accidental exactness in the vertex corrections of the self-energy
\cite{Atambo2026gw}, and extends our earlier \textit{ab initio} MBPT
applications of the same framework \cite{Atambo2019}.

Limitations: our $G_0$ is noninteracting, the clusters are one-dimensional,
and $v_q$ is a model interaction. These choices are deliberate: they isolate
the pole-representation error from all other approximation channels. The
diagnostic itself, being a property of $\varepsilon^{-1}$ alone, is
representation-independent and testable in \textit{ab initio} workflows.

\section{Conclusion}

Using an exact Lehmann-rank-one representation of the screened interaction
and an analytic $G_0W$ convolution, we have benchmarked the plasmon-pole
approximation without numerical frequency integration. The PPA is exact for
single-scale screening, fails systematically ($\sim1$-$2t$ gap errors) when
low-energy interband and high-energy Mott channels coexist, and can
artificially preserve quasiparticles that exact screening destroys. A
moment-based polydispersity of the loss function predicts the failure at
negligible cost. All code and data are available from the author.

\begin{acknowledgments}
The author acknowledges Kenya Education Network (KENET) research services for computing resources.
\end{acknowledgments}

\begin{figure}[htbp]
    \centering
    \includegraphics[width=\columnwidth]{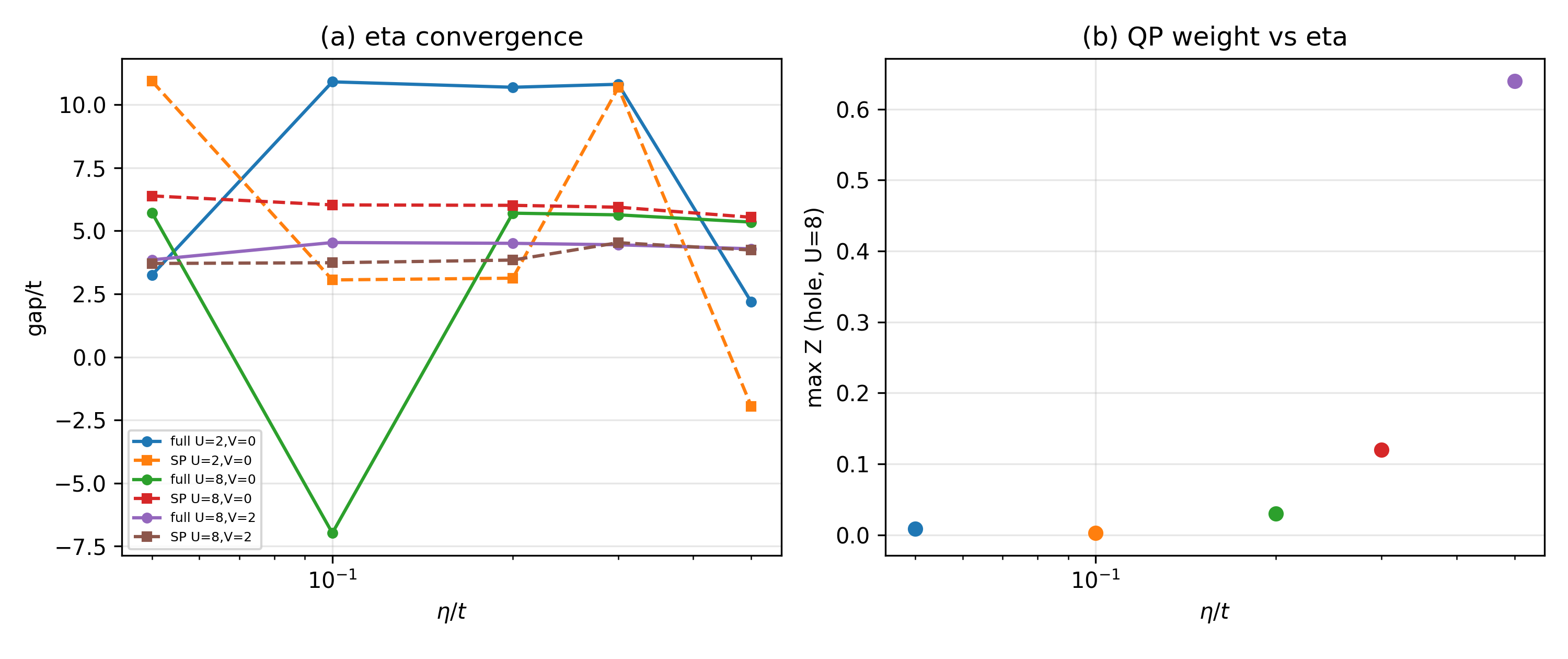}
    \caption{Thermodynamic-limit emulation via the boson lifetime parameter $\eta$. (a) Quasiparticle gaps for the exact (solid) and single-pole (dashed) representations, showing a stable plateau for strongly correlated states ($U=8$) at $\eta \ge 0.2t$. (b) The maximum quasiparticle weight $Z$ of the exact hole state at $U=8, V=0$, illustrating the fragmentation of the quasiparticle ($Z \to 0$) as $\eta \to 0$.}
    \label{fig:eta}
\end{figure}

\appendix

\section{Thermodynamic-limit emulation, $\eta$-convergence, and
quasiparticle fragmentation}
\label{app:eta}

A finite cluster yields a discrete set of undamped bosonic poles, whereas the
bulk screened interaction possesses a continuum with intrinsic damping. We
emulate the thermodynamic limit by assigning the Lehmann poles a lifetime
$\eta$ (Sec.~II) and treat $\eta$ as a controlled regulator. Three key observations emerge from the $\eta$-sweep (Table~\ref{tab:eta}, Fig.~\ref{fig:eta}):

(i) \emph{Quasiparticle fragmentation at small $\eta$.} As $\eta \to 0$, the exact $\mathrm{Re}\,\Sigma$
develops a forest of sharp poles. Consequently, the single-particle spectral weight is
fragmented into many incoherent shake-up lines-the finite-size precursor of the
bulk satellite continuum. For the strongly correlated configuration A ($U=8, V=0$), the maximum quasiparticle weight $Z$ on the hole branch is only $0.01$ at $\eta=0.05$ and vanishes ($Z \to 0$) at $\eta=0.10$. The collapse of the main-quasiparticle weight is
reminiscent of polaronic self-trapping, where $Z\to0$ at the
self-trapping crossover \cite{Holstein1959,Feynman1955,Devreese2009}. 
As $\eta$ increases, the poles smear into a continuum and a well-defined main quasiparticle emerges, with $Z$ rising to $0.12$ at $\eta=0.3$ and $0.64$ at $\eta=0.5$.
The static representation, lacking bosonic dynamics, trivially retains $Z=1$;
hence $\eta\to0$ is not a meaningful quasiparticle limit on finite clusters.

(ii) \emph{Gap plateau for correlated states.} For the strongly correlated configurations ($U=8$, $V=0,2$), the full and SP gaps exhibit a clear plateau for $\eta \ge 0.2$. For instance, the exact gap for configuration A varies smoothly from $5.70t$ ($\eta=0.2$) to $5.35t$ ($\eta=0.5$), while the SP gap tracks it from $6.01t$ to $5.54t$. The PPA error ($\sim 0.2 - 0.3t$) remains robust across this plateau. 

(iii) \emph{Sensitivity of weakly correlated states.} The weakly correlated control ($U=2, V=0$) exhibits large fluctuations in the extracted gap across $\eta$. This occurs because, at weak coupling, the exact self-energy possesses a dense manifold of low-energy satellite roots, and the root-finding algorithm hops between them as the broadening changes. This reinforces our central thesis: in the presence of complex, multi-pole structures, naive quasiparticle extraction is unstable, and PPA can artificially stabilize a single, sharp quasiparticle that does not physically exist in the exact spectrum.

We therefore quote all production results at
$\eta=0.3t$ (one band) and $\eta=0.2t$ (two band), representing the onset of the continuum plateau for strongly correlated states.

\begin{table}[b]
\caption{Quasiparticle gaps ($t$) vs boson lifetime $\eta$, $L=6$ one-band
ring, $k=0\to\pi$. ``max $Z$'' is the largest quasiparticle weight of the hole
branch at $U=8,V=0$, illustrating fragmentation at small $\eta$.}
\label{tab:eta}
\begin{ruledtabular}
\begin{tabular}{c|cc|cc|cc|c}
$\eta$ & ctrl full & ctrl SP & A full & A SP & B full & B SP & max $Z$ \\
\hline
0.05 & 3.25 & 10.94 & 5.72 & 6.39 & 3.85 & 3.71 & 0.01 \\
0.10 & 10.91 & 3.06 & -6.98 & 6.03 & 4.54 & 3.74 & 0.00 \\
0.20 & 10.70 & 3.13 & 5.70 & 6.01 & 4.51 & 3.84 & 0.03 \\
0.30 & 10.81 & 10.68 & 5.64 & 5.94 & 4.45 & 4.53 & 0.12 \\
0.50 & 2.18 & -1.95 & 5.35 & 5.54 & 4.29 & 4.24 & 0.64 \\
\end{tabular}
\end{ruledtabular}
\end{table}

All exact diagonalizations ($\dim\mathcal{H}=924$ and $1820$) and all
quasiparticle solves run in seconds per configuration on a single workstation
core; no high-performance computing was used at any stage of this work.

\section{Analytical machinery}
\label{app:math}

\subsection{Exact pole representation of $\varepsilon^{-1}$}
\label{app:rankone}

With $X\equiv\omega I-A$, $\mathbf{b}=(w_i,-w_i)^T$, $\mathbf{c}=\mathbf{1}$ and
$u(\omega)\equiv\mathbf{c}X^{-1}\mathbf{b}=\chi(q,\omega)$, the Woodbury
identity gives
$(X-v\mathbf{b}\mathbf{c})^{-1}=X^{-1}+X^{-1}\mathbf{b}\,v\,(1-u v)^{-1}\,
\mathbf{c}X^{-1}$, hence
\begin{equation}
1+v\,\mathbf{c}(\omega I-M)^{-1}\mathbf{b}
=1+vu+\frac{v^2u^2}{1-vu}=\frac{1}{1-vu}
=\varepsilon^{-1}(q,\omega),
\label{eq:appwoodbury}
\end{equation}
with $M=A+v\mathbf{b}\mathbf{c}$, proving Eq.~(\ref{eq:rankone})-(\ref{eq:Wpoles}). The poles $\{\lambda_m\}$
are the eigenvalues of $M$; with biorthogonal eigenvectors
$M r_m=\lambda_m r_m$, $l_m^{\dagger}M=\lambda_m l_m^{\dagger}$,
$l_m^{\dagger}r_n=\delta_{mn}$, the residues are
$R_m=v(\mathbf{c}\cdot r_m)(l_m^{\dagger}\cdot\mathbf{b})$, Eq.~(8). For
$\eta>0$ we verify $\mathrm{Im}\,\lambda_m<0$ for all $m$ (causality) and
reconstruct $\varepsilon^{-1}$ on a real-frequency grid with maximum deviation
$\le6\times10^{-9}$. The static limit is recovered exactly:
$\varepsilon^{-1}(q,0)=1/(1+2v m_{-1})$.

\subsection{Analytic $G_0W$ convolution}
\label{app:conv}

Inserting $W(q,\omega)=v_q+\sum_m R_m^{(q)}/(\omega-\lambda_m^{(q)})$ into the
zero-temperature $G_0W_0$ self-energy and performing the frequency integral
yields the standard quasiboson structure
\cite{Onida2002,Aryasetiawan1998}: bare exchange $-\sum_q v_q n_{k-q}$, an
\emph{emission} term for each positive-energy boson pole,
$(1-n_{k-q})R_m/(\omega-\varepsilon_{k-q}-\lambda_m)$, and an
\emph{absorption} term from the negative-frequency poles. Because the
negative-frequency poles of $\varepsilon^{-1}$ carry negative spectral weight
($R_m<0$ for $\mathrm{Re}\lambda_m<0$), the absorption term equals the
conventional $+n_{k-q}|g_m|^2/(\omega-\varepsilon_{k-q}+\Omega_m)$ with
$|g_m|^2=-R_m>0$. Collecting both channels gives Eq.~(\ref{eq:sigma}). For real, positive
residues Eq.~(10) reduces exactly to the familiar $g^2$ quasiparticle-boson
Hamiltonian form; our statement is therefore a strict generalization of it to
\emph{damped} (complex) bosons with exact residues. The quasiparticle weight
follows analytically,
$Z^{-1}=1-\partial_\omega\mathrm{Re}\Sigma
=1+\sum_{qm}\mathcal{W}_m^{(q)}\,\mathrm{Re}(\omega-\varepsilon_{k-q}
-\lambda_m^{(q)}+i0^+)^{-2}$,
used for branch selection (main quasiparticle $=$ largest positive $Z$).

\subsection{Properties of the single-pole model and of the polydispersity}
\label{app:sp}

For $\chi_{\rm SP}(\omega)=A_{\rm SP}[({\omega-\Omega_{\rm SP}+i\eta})^{-1}
-(\omega+\Omega_{\rm SP}+i\eta)^{-1}]$ with Eq.~(\ref{eq:sp}):
$\chi_{\rm SP}(0)=-2A_{\rm SP}/\Omega_{\rm SP}=-2m_{-1}=\chi(0)$ and
$\lim_{\omega\to\infty}\omega^2\chi_{\rm SP}=2A_{\rm SP}\Omega_{\rm SP}=2m_1$,
i.e. the static polarizability and the $f$-sum-rule moment are preserved
exactly. The zero of $\varepsilon_{\rm SP}$ is
$\omega_{\rm pl}^2=\Omega_{\rm SP}^2+2v_qA_{\rm SP}\Omega_{\rm SP}$.
The polydispersity $P_q=\sqrt{m_2m_0-m_1^2}/m_1$ is (a) zero iff the spectrum
is a single line (Cauchy-Schwarz equality), (b) invariant under rescaling of
the weights, and (c) monotonic in the separation of two lines at fixed weight
ratio; it therefore quantifies ``multi-pole content'' independently of
interaction strength. The aggregate $P=\sum_q P_q$ is computable in any
$GW$ code from three frequency moments of the loss function at a few momenta.

\end{document}